\documentclass[showpacs,showkeys,prd,nofootinbib]{revtex4}
\usepackage{amsmath}

\newcommand{\sumint}{\hbox{$\sum$}\!\!\!\!\!\!\!\int }

\begin{document}

\title{Gauge covariant heat kernel expansion at finite temperature}

\date{February, 2003}

\author{Stefan Leupold}


\affiliation{Institut f\"ur Theoretische Physik, Universit\"at
Giessen, D-35392 Giessen, Germany}

\begin{abstract}
The heat kernel method is extended to the case of finite temperature. Special emphasis
is given to the study of gauge theories. Due to the
compactness of space in the Euclidean time direction (inverse temperature) the field
strength cannot completely characterize the gauge fields. This is just a
manifestation of the Aharonov-Bohm effect. The field strength has to be supplemented
by the Polyakov loop. Only if the latter is taken into account one obtains gauge
covariant results for the generalized Seeley-DeWitt coefficients of the heat kernel 
expansion.
\end{abstract}

\maketitle

\section{Introduction}
\label{sec:intro}

Finite temperature quantum field theory has gained widespread applications in different
areas of physics
like e.g.~for the description of the early universe or the study of heavy-ion collisions.
This work is devoted to a finite temperature generalization of a non-perturbative
technique which is very well established for the vacuum case. 
The heat kernel expansion \cite{Schwinger:1951nm,dewitt,Seeley:1967ea}
(for a review see e.g.~\cite{Ball:1989xg})
is frequently used in the context of the path integral
method to integrate out degrees of freedom in a non-perturbative way. The outcome
is a derivative expansion of the fields which correspond to the remaining degrees of 
freedom which were not integrated out. In other words the heat kernel expansion
yields an effective field theory (cf.~e.g.~\cite{Ball:1989xg,espraf,gasleut1}). 
A typical application
which we also address here is to integrate out fermions from the action of a gauge
theory which is bilinear in the fermion fields. The obtained fermionic functional 
determinant is processed further by the heat kernel expansion. In this way the 
ultraviolet divergent terms which require renormalization can be properly isolated
e.g.~using the $\zeta$-function method \cite{Ball:1989xg}. It may appear that part of 
the symmetries of the original action are spoiled by the need for renormalization.
Such an appearance of anomalies can also be well accounted for with the heat kernel
method. In the following we shall not consider anomalies explicitly. We keep
the number of dimensions as well as the gauge group completely general. We restrict
ourselves, however, to a flat space-time structure.

Naively generalizing
the heat kernel expansion to finite temperature we will encounter problems concerning
the gauge symmetries as we will see below. 
This, however, is not a manifestation of an anomaly, as it
appears for any gauge symmetry and only at finite temperature. We will solve that
problem subsequently by presenting an improved heat kernel expansion which preserves
the gauge symmetries (apart from potential anomalies of course). Throughout this
work we will use the Matsubara formalism \cite{Matsubara:1955ws,Das:1997gg} 
to represent fields at 
finite temperature. It amounts to a Euclidean metric with the boson (fermion) fields
obeying (anti-)periodic boundary conditions in the Euclidean time direction with
periodicity $\beta=1/T$ where $T$ is the temperature of the heat bath the fields are
coupled to. 

The task to extend the heat kernel expansion to finite temperature has
already been addressed in \cite{Boschi-Filho:1992ah}. It has been claimed there that the ratio
of the finite temperature heat kernel and the vacuum one is just a temperature
dependent constant independent of the considered fields. In contradiction to that
statement the authors of \cite{Actor:1998cn} presented a counter-example where the mentioned
ratio does depend on the gauge fields.
It is the purpose of the present work to clarify this contradiction and to present
a valid generalization of the heat kernel expansion to the finite temperature case.
In vacuum the heat kernel expansion is an expansion in powers of the field
amplitudes and their derivatives. For gauge fields this amounts to an expansion
in powers of the field strengths and their gauge covariant derivatives. All the
other information present in the vector potentials is due to the gauge freedom and
therefore of no physical significance. 
At finite temperature, however, the field strengths and their derivatives are no longer 
sufficient to characterize the gauge fields. The crucial point is that due to the
periodicity requirement mentioned above the Euclidean time direction is compactified.
This leads to the appearance of an Aharonov-Bohm effect \cite{Aharonov:1959fk}. 
The Polyakov loop 
\cite{Polyakov:1978vu} supplements the information contained in the field strengths. 
It is defined by
\begin{equation}
  \label{eq:defpoly}
L(x) := P \exp 
\left[-\int\limits_{x_0}^{x_0+\beta} dt \, A_0(t,\vec x) \right] 
\end{equation}
where $A_0$ is the zeroth component of the gauge potential and $P$ denotes path 
ordering. The latter is required for non-abelian gauge theories
to obtain a gauge covariant result. Obviously the Polyakov loop constitutes a
non-local object in terms of the gauge fields. It is easy to check that a
Taylor expansion of $L$ in powers of the gauge potential and its derivatives e.g.~around
the point $x$ would yield objects which are ill-defined with respect to gauge
transformations. Therefore, the Polyakov loop should be treated on equal footing
with the field strength without any further expansion in terms of the amplitudes of
the gauge potential. On the other hand, a heat kernel expansion is just an expansion
in terms of the field amplitudes. Therefore to prevent an erroneous expansion of 
the Polyakov loop, the latter has to be isolated right from the start to obtain
a heat kernel expansion which is appropriate for the finite temperature case.

The paper is organized in the following way: In the next section we take
the usual method to perform a heat kernel expansion and naively 
generalize it to the finite
temperature case. To some extent we follow here the work of \cite{Boschi-Filho:1992ah}. At a certain
stage, however, we have to point out differences between our findings and the ones of
\cite{Boschi-Filho:1992ah}. Finally we have to face a problem concerning the gauge covariance of the
results obtained with this naive generalization of the heat kernel expansion technique.
In Section \ref{sec:covheat} we present an improved method to perform the heat kernel
expansion at finite temperature preserving gauge covariance in every single step. 
Gauge covariant Seeley-DeWitt coefficients are calculated in Section \ref{sec:gvsdw}.
It is also shown in this section that the results reduce to the already known results
for the vacuum case. In addition, the general results for the Seeley-DeWitt 
coefficients are applied to the case of two-dimensional QED with constant fields.
For this special case the heat kernel and thus also all its Seeley-DeWitt coefficients
can be calculated exactly \cite{Actor:1998cn}. This provides the possibility to cross-check
our results with the ones presented in \cite{Actor:1998cn}. Finally Section \ref{sec:sum}
summarizes our findings. Three Appendices are added containing technical aspects
of the presented formalism.

\section{Naive representation for the heat kernel}
\label{sec:naiv}

Consider the differential operator in Euclidean space
\begin{equation}
  \label{eq:deffop}
\Delta_x = -D^x_\mu D^x_\mu +X(x)
\end{equation}
with the gauge covariant derivative
\begin{equation}
  \label{eq:gaugeder}
D^x_\mu = \partial^x_\mu + A_\mu(x)  \,.
\end{equation}
Here the functions $A_\mu$ and $X$ are no operators (but might be matrix-valued due to 
the group structure of the gauge group under consideration). 
Under gauge transformations the introduced quantities transform as
\begin{equation}
  \label{eq:gaugetrafo}
X(x) \to V(x)\, X(x) \, V^{-1}(x) \,, \qquad D^x_\mu \to V(x)\, D^x_\mu V^{-1}(x) \,.
\end{equation}
The heat kernel is given by
\begin{equation}
  \label{eq:naiveheatk}
H(x,x';\tau) := \left( x \vert e^{-\tau \Delta} \vert x' \right) 
= e^{-\tau \Delta_x} \, \delta(x-x')  \,.
\end{equation}
We shall assume that the differential operator (\ref{eq:deffop}) originates from
a fermionic determinant which we want to evaluate by the heat kernel technique 
\cite{Ball:1989xg}. A generalization of the presented formalism to the bosonic case
is in principle straightforward.
The representation for the heat kernel on the right hand side of (\ref{eq:naiveheatk})
is completely appropriate for the vacuum case.
In this section we will generalize this heat kernel to the finite temperature case.
We will point out the problem which appears when this heat kernel is used at finite
temperature. In the next section we will present an improved representation for
the heat kernel which is better suited for the treatment at finite temperature.
Hence some of the formulae which we present in this section should not be regarded as
valuable results but as illustrations of the problems which appear at finite temperature.

For finite $T$ the $\delta$-function which appears in (\ref{eq:naiveheatk}) is given
by
\begin{equation}
  \label{eq:defdeltafint}
\delta(x-x') = \sum\limits_{n = -\infty}^{+\infty} 
T \int\! {d^{D-1}k \over (2 \pi)^{D-1}} \, e^{i \vec k (\vec x - \vec x')}
\, e^{i \omega_n \, (x_0 - x_0')} 
=: \sumint  {d^{D}k \over (2 \pi)^{D}} \, e^{i k \cdot (x - x')}
\end{equation}
where $D$ is the number of space-time dimensions.
Being interested in the calculation of a fermionic determinant via the heat kernel
method the discrete energies appearing in (\ref{eq:defdeltafint}) are defined as
\cite{Matsubara:1955ws,Das:1997gg}
\begin{equation}
  \label{eq:defenerg}
\omega_n = 2 \pi T \,\left( n + {1 \over 2} \right)   \,.
\end{equation}
It is straightforward to show that the diagonal entries of $H$ can be written as
\cite{Nepomechie:1985wt}
\begin{equation}
  \label{eq:expanhs}
H(x,x;\tau) = \sumint {d^{D}k \over (2 \pi)^{D}} \, e^{-\tau \, (k^2+M^2)}
\left. 
\sum\limits_{m=0}^\infty {1 \over m!} \, \tau^m \, (2 i k \cdot D_x - \Delta_x + M^2)^m
f(x) \right\vert_{f = 1}  \,.
\end{equation}
We have introduced an arbitrary mass term $M$ to keep the formalism as general as
possible. All derivatives in (\ref{eq:expanhs}) act on an arbitrary function $f$ which
has to be put to 1 at the very end. We want to emphasize already here that the precise
meaning of ``covariant derivative acting on $f$'' is important. As $f$ basically
substitutes a fermionic field the meaning of $D^x_\mu f(x)$ really is
$\partial_\mu^x f(x)+ A_\mu(x)\,f(x)$ and {\em not} $[D_\mu^x,f(x)]$. Therefore
eventually putting $f$ to 1, a term like $A_\mu(x)$ might remain! This is in contrast
to the claim in \cite{Boschi-Filho:1992ah} that $D_\mu f$-terms should be dropped in the
end (cf.~the statement between eqs.~(3.17) and (3.18) in \cite{Boschi-Filho:1992ah}). This will
have important consequences as we will see below.

Next we introduce generalized Seeley-DeWitt coefficients 
\cite{Seeley:1967ea,dewitt,Nepomechie:1985wt,Ball:1989xg} as follows:
\begin{equation}
  \label{eq:seeley}
H(x,x;\tau) = \left( {1 \over 4 \pi \tau} \right)^{D/2} e^{-\tau M^2}
\sum\limits_{n =0}^\infty g_n(x;\sqrt{\tau} T) \, \tau^n  \,.
\end{equation}
Note that at finite temperature there is a remaining $\tau$ dependence of the 
coefficients $g_n$ but only via the specific combination $\sqrt{\tau} T$ \cite{Boschi-Filho:1992ah}. 
Hence the $g_n$'s
become independent of $\tau$ for the vacuum case $T=0$. 
To calculate e.g.~the first two coefficients, $g_0$ and $g_1$, we have to evaluate the 
sum over $m$
in (\ref{eq:expanhs}) up to $m=2$.
We note that all terms with odd numbers of
$k_\mu$ drop out by the $k$ summation/integration. We first calculate
\begin{equation}
  \label{eq:defcm}
c_m := \left. \left . (2 i k \cdot D_x - \Delta_x + M^2)^m f(x) \right\vert_{f = 1} \,\,
\right\vert_{\mbox{drop odd powers of $k$}}  \,\,.
\end{equation}
We get
\begin{subequations}
\label{eq:ccoeff}
\begin{eqnarray}
\label{eq:ccoeff0}
c_0 & = & 1 \,,
\\  
\label{eq:ccoeff1}
c_1 & = & D_\mu A_\mu - \tilde X  \,,
\\
\label{eq:ccoeff2}
c_2 & = & -4 k_\mu k_\nu D_\mu A_\nu + D_\nu D_\nu D_\mu A_\mu - D_\mu D_\mu \tilde X
- \tilde X D_\mu A_\mu + \tilde X^2  \,,
\end{eqnarray}
\end{subequations}
where we have introduced 
\begin{equation}
  \label{eq:defxtilde}
\tilde X := X - M^2  \,.
\end{equation}
Note that the momenta scale as 
$1/\sqrt{\tau}$ due to the $\exp(-\tau k^2)$ term in (\ref{eq:expanhs}). Therefore 
it is easy to see that $c_0$ contributes to $g_0$, $c_1$ and the first term of $c_2$
contribute to $g_1$, and the remaining terms of $c_2$ contribute to $g_2$. Hence
\begin{subequations}
    \label{eq:acoeff}
  \begin{eqnarray}
g_0 & = & (4 \pi \tau)^{D/2} \sumint {d^{D}k \over (2 \pi)^{D}} \, e^{-\tau k^2} \, c_0
= I_0(\sqrt{\tau}T)  \,,
\\
g_1 & = & (4 \pi \tau)^{D/2} \sumint {d^{D}k \over (2 \pi)^{D}} \, e^{-\tau k^2} \, 
\left(c_1 - {1 \over 2} \, \tau \, 4 k_\mu k_\nu D_\mu A_\nu \right)
= (D_\mu A_\mu -\tilde X) \, I_0(\sqrt{\tau}T) - D_0 A_0 \, I_{2t}(\sqrt{\tau}T)  
- D_i A_i \, I_{2s}(\sqrt{\tau}T)
\nonumber \\
    \label{eq:acoeffb}
& = & -\tilde X \, I_0(\sqrt{\tau}T) + D_0 A_0 \, 
\left( I_0(\sqrt{\tau}T) - I_{2t}(\sqrt{\tau}T) \right)
\,,
  \end{eqnarray}
\end{subequations}
where the relevant sum-integrals are collected in Appendix \ref{sec:app}.

The important point now is that the term $D_0 A_0 $ does not transform homogeneously
under gauge transformations in contrast to $\tilde X$.\footnote{Note that $\tilde X$ 
transforms in the same way as $X$.} Therefore
this cannot be a proper result for the generalized Seeley-DeWitt coefficient. This
problem does not show up at zero temperature since then both integrals $I_0$ and
$I_{2t}$ become identical (cf.~Appendix \ref{sec:app}). 
Hence the dubious term drops out in this case. We note
that this $D_0 A_0$-term was overlooked in \cite{Boschi-Filho:1992ah}. It was still present in
eq.~(3.15) of \cite{Boschi-Filho:1992ah} but was disregarded afterwards by arguing that $D_0 f$
vanishes in the end. As already pointed out, this would be true for $[D_0,f]$ but
not for $D_0 f$. Ignoring this subtle difference the authors of \cite{Boschi-Filho:1992ah} did
not encounter any problem concerning gauge invariance. However, as shown above there is 
a problem with gauge invariance and actually there should be one on physical grounds:
By construction the heat kernel expansion (\ref{eq:expanhs}) is an expansion
in powers of field amplitudes and their derivatives. Any non-local object is 
expanded in terms of local ones. In vacuum this is a procedure which preserves gauge 
invariance --- apart from well-defined anomalies which should be there on physical
grounds. The reason why this procedure works for the vacuum case can be traced back
to the fact that the field strengths (and their derivatives) are sufficient to 
characterize the physical content of the gauge fields. The field strengths, however,
are local objects. The situation is completely different at finite temperature
as already mentioned in the introduction. In this case the Polyakov loop carries
additional information. This object is manifestly non-local. An expansion of the
Polyakov loop in powers of the field amplitudes would yield misleading gauge dependent
results. Hence the Polyakov loop has to be isolated as an additional degree of freedom
{\em before} processing the heat kernel expansion.

Having established the physical reason for the erroneous result (\ref{eq:acoeffb})
we next have to analyze what went wrong on the technical level when proceeding from
(\ref{eq:naiveheatk}) to (\ref{eq:expanhs}). Concerning the behavior with respect
to gauge transformations we observe that formally the heat kernel (\ref{eq:naiveheatk}) 
transforms as a non-local object,
\begin{equation}
  \label{eq:trafoheat}
H(x,x';\tau) \to V(x)\, e^{-\tau \Delta_x} \, V^{-1}(x)\, \delta(x-x') =
V(x)\, e^{-\tau \Delta_x} \, \delta(x-x') \, V^{-1}(x') =
V(x)\, H(x,x';\tau)  \, V^{-1}(x')  \,,
\end{equation}
on account of the transformation properties (\ref{eq:gaugetrafo}). It is illustrative
to decompose this transformation property in the following way:
\begin{equation}
  \label{eq:decompopdelta}
e^{-\tau \Delta_x} \to V(x)\, e^{-\tau \Delta_x} \, V^{-1}(x) \quad \mbox{and}
\quad \delta(x-x') \to V(x)\, \delta(x-x')  \, V^{-1}(x') \,.
\end{equation}
Correspondingly the left hand side of (\ref{eq:expanhs})
transforms gauge covariantly
\begin{equation}
  \label{eq:trafoheatdiag}
H(x,x;\tau) \to V(x)\, H(x,x;\tau)  \, V^{-1}(x)   \,.
\end{equation}
In contrast, the single Fourier modes on the right hand side of 
(\ref{eq:expanhs}) do not show that behavior with respect to gauge transformations.
In general, only after summing up all Fourier modes and performing the sum over $m$ one 
obtains the heat kernel with its well-defined transformation properties. For the
vacuum case it turns out that already the summation over the Fourier modes is sufficient;
each single Seeley-DeWitt coefficient obtained in that way transforms gauge covariantly.
As we have demonstrated above this is no longer true at finite temperature. Therefore
it is more appropriate to use an approach which preserves in each step as much of the 
good transformation properties as possible. To be specific, we are looking for a 
Fourier type
representation of the $\delta$-function which preserves the transformation property
(\ref{eq:decompopdelta}) for each single Fourier mode. In this way we will never 
encounter the problem that
only a sum of objects transforms properly while the single terms do not. We stress
once more that such an approach is not mandatory for the vacuum case. After the 
summation of the Fourier modes is performed the results transform properly with respect
to gauge transformations.

\section{Gauge covariant heat kernel}
\label{sec:covheat}

To obtain gauge covariant results for the Seeley-DeWitt coefficients it might be
not mandatory but it is obviously
more save to start with a Fourier type representation of the $\delta$-function
which transforms gauge covariantly in every single Fourier mode. An important ingredient
for such an approach is the gauge link 
(see e.g.~\cite{Yang:1974kj,Bralic:1980ra,Elze:1986qd})
\begin{equation}
  \label{eq:defU}
U[A](x;x') := P \exp\left[ - \int\limits_{x'}^x \!\! dz_\mu A_\mu(z) \right]
\end{equation}
where $P$ denotes path ordering. For our purposes it is useful to choose the following
path parametrized by $z$: the path connects with straight lines the sequence of points 
$x'=(x'_0,\vec x') \to (\tau_0,\vec x') \to (\tau_0, \vec x) \to x=(x_0,\vec x)$.
Here $\tau_0$ is an arbitrary time. It will turn out that none of the final results 
will depend on $\tau_0$. From the technical point of view this aspect can be seen as 
a consistency check of the whole approach. 
For the chosen path we find
\begin{equation}
  \label{eq:U1}
U[A](x,x) = 1  \,.
\end{equation}
Another important property of gauge links is their additivity 
\cite{Yang:1974kj,Bralic:1980ra,Elze:1986qd}:
\begin{equation}
  \label{eq:Uadd}
U[A](x;x') \, U[A](x';x'') = U[A](x;x'')
\end{equation}
valid for arbitrary $x'$ on the contour connecting $x$ and $x''$. 
Finally under gauge transformations $U$ transforms as 
\cite{Yang:1974kj,Bralic:1980ra,Elze:1986qd}
\begin{equation}
  \label{eq:Utrafo}
U[A](x;x') \to V(x)\, U[A](x;x')  \, V^{-1}(x')  \,.
\end{equation}
Therefore it is tempting to write
\begin{equation}
  \label{eq:deltaprelim}
\delta(x-x') = \delta(x-x') U[A](x,x') 
= \sumint  {d^{D}k \over (2 \pi)^{D}} \, e^{i k \cdot (x - x')} \, U[A](x,x') \,.
\end{equation}
Indeed, now the modified single Fourier modes transform gauge covariantly, i.e.~in the 
same way as the $\delta$-function in (\ref{eq:decompopdelta}):
\begin{equation}
  \label{eq:trafomodfourier}
e^{i k \cdot (x - x')} \, U[A](x,x') \to 
V(x)\, e^{i k \cdot (x - x')} \, U[A](x;x')  \, V^{-1}(x')  \,.
\end{equation}
However this cannot be the full story: At finite temperature the $\delta$-function
which corresponds to fermions must be antiperiodic in Euclidean time with periodicity 
$\beta = 1/T$:
\begin{equation}
  \label{eq:antiperdelta}
\delta(x_0 + \beta - x_0') = - \delta(x_0 - x_0')  \,.
\end{equation}
This is also true for the single Fourier modes 
\begin{equation}
  \label{eq:antiperefnc}
e^{i k_0 \, (x_0 + \beta - x_0')} = - e^{i k_0 \, (x_0 - x_0')}
\end{equation}
on account of (\ref{eq:defenerg}).
The modified Fourier modes displayed on the left hand side of (\ref{eq:trafomodfourier})
do not share that important feature as the gauge link $U$ is {\em not} 
periodic in time:
\begin{equation}
  \label{eq:Unonper}
U[A](x_0+\beta,\vec x;x_0',\vec x') 
= \underbrace{U[A](x_0+\beta,\vec x;x_0,\vec x)}_{=L(x)} 
\, U[A](x_0,\vec x;x_0',\vec x') \neq U[A](x_0,\vec x;x_0',\vec x') 
\end{equation}
where we have used (\ref{eq:Uadd}). Note that the term which spoils the periodicity
is nothing but the Polyakov loop defined in (\ref{eq:defpoly}).

We now construct a modified
gauge link which both has the transformation property (\ref{eq:Utrafo}) and is periodic 
in time. The key point is to isolate the
(generalized) zero mode of the time component of the gauge potential which is closely
related to the Polyakov loop. We first define the generalized zero mode
\begin{equation}
  \label{eq:defa0}
a_0(x) := -T \log L(x) 
= -T \log U[A](x_0+\beta,\vec x;x_0,\vec x)   \,.
\end{equation}
The properties of $a_0$ are collected in Appendix \ref{sec:a0prop}. In general
already the log-function of complex numbers is multiple valued. This is of course
also true for the log-function of matrices.
In the following we use the log-function defined
via its Taylor expansion (cf.~Appendix \ref{sec:a0prop})
\begin{equation}
  \label{eq:deflog}
\log (1+{\cal O}) := 
\sum\limits_{n=1}^\infty (-1)^{n+1} \, {1 \over n} \, {\cal O}^n   \,.
\end{equation}
Hence we restrict ourselves to Polyakov loops with values close to the unity matrix.
We postpone the further discussion of that
issue to Section \ref{sec:sum}. We introduce the modified gauge link
\begin{equation}
  \label{eq:modgaugeconn}
\tilde U[A](x;x') := e^{x_0 \, a_0(x)} \, U[A](x;x') \, e^{-x_0' \, a_0(x')}  \,.
\end{equation}
Using the properties of $a_0$ collected in Appendix \ref{sec:a0prop}
it is straightforward to prove the following properties of $\tilde U$:
\begin{subequations}
\begin{eqnarray}
  \label{eq:propUtilde1}
\tilde U[A](x;x) & = & 1 \,,
\\
\tilde U[A](x_0+\beta,\vec x;x_0',\vec x') & = & \tilde U[A](x_0,\vec x;x_0',\vec x') 
\,, \\
\tilde U[A](x;x') & \to & V(x)\, \tilde U[A](x;x')  \, V^{-1}(x')  \,.
\end{eqnarray}
\end{subequations}
With these developments at hand the heat kernel can be represented as
\begin{equation}
  \label{eq:heatgaugecov}
H(x,x';\tau) = e^{-\tau \Delta_x} \left( \delta(x-x') \, \tilde U[A](x;x') \right) \,.
\end{equation}
For the diagonal part one gets
\begin{equation}
  \label{eq:diagheatgaugecov}
H(x,x;\tau) = \sumint {d^{D}k \over (2 \pi)^{D}} \, e^{-\tau \, (k^2+M^2)}
\left. 
\sum\limits_{m=0}^\infty {1 \over m!} \, \tau^m \, (2 i k \cdot D_x - \Delta_x + M^2)^m
\, \tilde U[A](x;x') \, \right\vert_{x' = x}  \,.
\end{equation}


\section{Gauge covariant Seeley-DeWitt coefficients}
\label{sec:gvsdw}

From the representation (\ref{eq:diagheatgaugecov}) it is straightforward to 
calculate the generalized Seeley-DeWitt coefficients defined in (\ref{eq:seeley}).
We present in the following the calculation of the first three generalized Seeley-DeWitt
coefficients. For this task we introduce
\begin{equation}
  \label{eq:deftildec}
\tilde c_m := \left. \left . (2 i k \cdot D_x - \Delta_x + M^2)^m
\, \tilde U[A](x;x') \, \right\vert_{x' = x}  \,\,
\right\vert_{\mbox{drop odd powers of $k$}}  \,\,.
\end{equation}
Note that by construction every $\tilde c_m$ transforms gauge covariantly in contrast to
$c_m$ as defined in (\ref{eq:defcm}).
For the lowest order terms we get
\begin{subequations}
\label{eq:ctcoeff}
\begin{eqnarray}
\label{eq:ctcoeff0}
\tilde c_0 & = & 1 \,,
\\  
\label{eq:ctcoeff1}
\tilde c_1 & = & \left. (D_\mu D_\mu - \tilde X) \, \tilde U[A](x;x') 
\, \right\vert_{x' = x}  \,,
\\
\label{eq:ctcoeff2}
\tilde c_2 & = & \left. (-4 k_\mu k_\nu D_\mu D_\nu + D_\mu D_\mu D_\nu D_\nu 
- \tilde X D_\mu D_\mu - D_\mu D_\mu \tilde X + \tilde X^2 ) \,
\tilde U[A](x;x') \, \right\vert_{x' = x}  \,,
\\
\tilde c_3 & = & \left. 4 k_\mu k_\nu \, (-D_\mu D_\nu D_\alpha D_\alpha 
- D_\mu D_\alpha D_\alpha D_\nu - D_\alpha D_\alpha D_\mu D_\nu 
+ D_\mu D_\nu \tilde X + D_\mu \tilde X D_\nu + \tilde X D_\mu D_\nu )
\, \tilde U[A](x;x') \, \right\vert_{x' = x}  
\nonumber \\
\label{eq:ctcoeff3}
&& + \mbox{terms which contribute to $g_3$}  \,,
\\
\label{eq:ctcoeff4}
\tilde c_4 & = & \left. 16 k_\mu k_\nu k_\alpha k_\beta 
D_\mu D_\nu D_\alpha D_\beta \tilde U[A](x;x') \, \right\vert_{x' = x}  
+ \mbox{terms which contribute to $g_3$, $g_4$}  \,.
\end{eqnarray}
\end{subequations}
It is lengthy but straightforward to determine the corresponding generalized 
Seeley-DeWitt coefficients. 
In the course of the calculation there appears the following type of $n$-fold 
derivatives:
\begin{equation}
  \label{eq:timeder}
\left. D_{x,0}^n \, \tilde U[A](x;x') \, \right\vert_{x' = x} = a_0^n(x)   \,.
\end{equation}
The proof for this relation is given in Appendix \ref{sec:proof}.
The formulae collected in Appendix \ref{sec:app} serve to perform the $k_i$ integrations
and to rewrite the finite temperature sums.
The lowest order generalized Seeley-DeWitt coefficients turn out to be
\begin{subequations}
    \label{eq:acoeffcov}
\begin{eqnarray}
g_0 & = & I_0(\sqrt{\tau}T)  \,,
\\
g_1 & = & -\tilde X \, I_0(\sqrt{\tau}T) + a_0^2 \, 
\left( I_0(\sqrt{\tau}T) - I_{2t}(\sqrt{\tau}T) \right)
\,,
\\
g_2 & = & {1 \over 2} \, \tilde X^2 I_0(\sqrt{\tau}T) 
- {1 \over 6} \left[ D_i,[D_i,\tilde X] \right] I_0(\sqrt{\tau}T) 
- {1 \over 6} \left[ D_0,[D_0,\tilde X] \right] 
  \left( 3 I_0(\sqrt{\tau}T) - 2 I_{2t}(\sqrt{\tau}T) \right)
\nonumber \\ && {}
- \left( [D_0,\tilde X] \, a_0 + \tilde X a_0^2 \right)
  \left( I_0(\sqrt{\tau}T) - I_{2t}(\sqrt{\tau}T) \right)
+ {1 \over 12} [D_i,D_j]^2 I_0(\sqrt{\tau}T)
+ {1 \over 6} [D_i,D_0]^2 \, \left( 2 I_0(\sqrt{\tau}T) - I_{2t}(\sqrt{\tau}T) \right)
\nonumber \\ && {}
+ {1 \over 6} \left[ D_i, \left[ D_0, [D_i,D_0] \right] \right] \,
  \left( I_0(\sqrt{\tau}T) - I_{2t}(\sqrt{\tau}T) \right)
+ {1 \over 3} \left[ D_i, [D_i,D_0] \right] \, a_0 \, 
  \left( I_0(\sqrt{\tau}T) - I_{2t}(\sqrt{\tau}T) \right)
\nonumber \\ \label{eq:acoeffcov2}
&& {}
+ {1 \over 2} \, a_0^4 \, 
  \left( I_0(\sqrt{\tau}T) - 2 I_{2t}(\sqrt{\tau}T) + I_{4t}(\sqrt{\tau}T) \right)
\,.
\end{eqnarray}
\end{subequations}
This constitutes the main result of the present work. Of course, by construction
every coefficient transforms gauge covariantly.
Obviously the method can be
extended to calculate higher order coefficients. The expressions for $I_{\dots}$
are collected in Appendix \ref{sec:app}. Note that the coefficients $g_n$ do not
depend on the number of space-time dimensions $D$. The dependence of $g_n$ 
on the temperature
is non-analytic as can be seen by inspecting (\ref{eq:i0fin}) and (\ref{eq:i0combi}).
This leads to a very smooth transition of the vacuum to finite temperature results.
In the following we shall apply these general results to two special cases. The
purpose is to check the consistency of our results.

In vacuum all $I_{\dots}$-functions become unity (see Appendix \ref{sec:app}).
One obtains the standard results for this limiting case \cite{Ball:1989xg}:
\begin{subequations}
    \label{eq:acoeffcovvac}
\begin{eqnarray}
g^{\rm vac}_0 & = & 1  \,,
\\
g^{\rm vac}_1 & = & -\tilde X \,,
\\
g^{\rm vac}_2 & = & {1 \over 2} \, \tilde X^2 
- {1 \over 6} \left[ D_i,[D_i,\tilde X] \right] 
- {1 \over 6} \left[ D_0,[D_0,\tilde X] \right] 
+ {1 \over 12} [D_i,D_j]^2 
+ {1 \over 6} [D_i,D_0]^2 
\nonumber \\
& = & {1 \over 2} \, \tilde X^2 
- {1 \over 6} \left[ D_\mu,[D_\mu,\tilde X] \right] 
+ {1 \over 12} [D_\mu,D_\nu]^2
\,.
\end{eqnarray}
\end{subequations}

As a further check of our results (\ref{eq:acoeffcov}) we compare them to the special 
case discussed
in \cite{Actor:1998cn}: two-dimensional massless QED with a constant electric field strength
$E$ and in addition a constant zero mode. The (trace of the) heat kernel for such a 
system is calculated
exactly in \cite{Actor:1998cn}: 
\begin{equation}
  \label{eq:actorheat}
{\rm tr}H(x,x;\tau)= {E \over 2\pi} {1 \over \tanh E\tau} 
\left\{ 1 + 2 \sum\limits_{n=1}^\infty (-1)^n 
        \cos\left[ 
               n\, {E x_1 + 2 \pi T a \over T}
            \right]
        \exp \left(- {n^2 E \over 4 T^2 \tanh E\tau} \right)
\right\}    \,.
\end{equation}
Below we will expand the exact result to obtain
the corresponding generalized Seeley-DeWitt coefficients. The latter can be compared to 
our results which we will apply now to the case at hand.

The vector potentials are given by
\begin{equation}
  \label{eq:actorvpot}
A_0 = -i \, (E x_1 + 2 \pi T a) \,, \qquad A_1 = 0  \,.
\end{equation}
For the zero mode defined in (\ref{eq:defa0}) we find\footnote{The already mentioned
ambiguity concerning the logarithm will be discussed below.}
\begin{equation}
  \label{eq:twodimzeromode}
a_0 = A_0 = -i \, (E x_1 + 2 \pi T a) \,.
\end{equation}
The task is to calculate (approximately) the heat kernel for the operator
\begin{equation}
  \label{eq:twodimop}
\not \! \! D^2 = D_\mu D_\mu - X
\end{equation}
with
\begin{equation}
  \label{eq:twodimX}
X = -{1 \over 4} [\gamma_\mu,\gamma_\nu] [D_\mu,D_\nu] = -E \sigma_3
\end{equation}
where we have used the convention \cite{Actor:1998cn}
\begin{equation}
  \label{eq:convgamma}
\gamma_0 = -\sigma_1 \,, \qquad \gamma_1 = \sigma_2
\end{equation}
with the Pauli matrices $\sigma_i$.

For this case the generalized Seeley-DeWitt coefficients are (of course we use $M=0$):
\begin{subequations}
    \label{eq:acoefftwodim}
\begin{eqnarray}
g_0 & = & I_0(\sqrt{\tau}T)  \,,
\\
g_1 & = & E \sigma_3 \, I_0(\sqrt{\tau}T) - (E x_1+2\pi T a)^2 \, 
\left( I_0(\sqrt{\tau}T) - I_{2t}(\sqrt{\tau}T) \right)
\,,
\\
g_2 & = & 
- E \sigma_3 \, (E x_1 + 2 \pi T a)^2  \,
  \left( I_0(\sqrt{\tau}T) - I_{2t}(\sqrt{\tau}T) \right)
+ {1 \over 6} \, E^2 \, \left( I_0(\sqrt{\tau}T) + I_{2t}(\sqrt{\tau}T) \right)
\nonumber \\ && {}
+ {1 \over 2} \, (E x_1 + 2\pi T a)^4 \, 
  \left( I_0(\sqrt{\tau}T) - 2 I_{2t}(\sqrt{\tau}T) + I_{4t}(\sqrt{\tau}T) \right)
\,.
\end{eqnarray}
\end{subequations}

The expansion of (\ref{eq:actorheat}) has to be performed in the following way:
\underline{Expand in powers of $\tau$, but with $T^2\tau$ fixed.} 
There is, however, one exception
to this rule, namely the $2\pi T a$ term. The temperature appearing there does not
originate from integrating out the thermal fermions but was put in
by hand in \cite{Actor:1998cn} to make $a$ dimensionless. To account for that fact we use
(\ref{eq:twodimzeromode}) and rewrite (\ref{eq:actorheat}) as
\begin{equation}
  \label{eq:actorheat2}
{\rm tr}H(x,x;\tau)= {1 \over 2\pi\tau} {E\tau \over \tanh E\tau} 
\left\{ 1 + 2 \sum\limits_{n=1}^\infty (-1)^n 
        \cos\left[ 
               \sqrt{\tau} \, n \, {i a_0 \over \sqrt{\tau}T}
            \right]
        \exp \left(- {n^2 \over 4 \tau T^2} {E \tau \over \tanh E\tau} \right)
\right\}    \,.
\end{equation}
Now we apply our expansion rule to (\ref{eq:actorheat2}):
\begin{eqnarray}
  \label{eq:expactorheat}
{\rm tr}H(x,x;\tau) & \approx &
{1 \over 2\pi\tau} \, \left( 1 + {1 \over 3} \tau^2 E^2 \right)
\nonumber \\ && \times
\left\{ 1 + 2 \sum\limits_{n=1}^\infty (-1)^n 
     \left( 1 + \tau \, {n^2 a_0^2 \over 2 \tau T^2} 
            + \tau^2 \, {n^4 a_0^4 \over 24 \tau^2 T^4} \right)
     \left( 1- {1 \over 3} \tau^2 E^2 {n^2 \over 4 \tau T^2}
     \right)
     \exp \left(- {n^2 \over 4 \tau T^2} \right)
\right\} 
\nonumber \\
& \approx & {1 \over 2\pi\tau} \, \left( 1 + {1 \over 3} \tau^2 E^2 \right)
\left\{
   I_0 + \left( \tau a_0^2 - {1 \over 6} \tau^2 E^2 \right) (I_0 - I_{2t})
   + {1 \over 2} \tau^2 a_0^4 \, (I_0 - 2 I_{2t} + I_{4t} )
\right\}
\nonumber \\
& \approx & {1 \over 2\pi\tau} \, 
\left[
   \left( 1 + {1 \over 3} \tau^2 E^2 \right) I_0 
   + \left( \tau a_0^2 - {1 \over 6} \tau^2 E^2 \right) (I_0 - I_{2t})
   + {1 \over 2} \tau^2 a_0^4 \, (I_0 - 2 I_{2t} + I_{4t} )
\right]
\end{eqnarray}
where we have used the relations (\ref{eq:i0fin}) and (\ref{eq:i0combi}) to rewrite
the finite temperature sums. Finally recalling (\ref{eq:seeley}) we can read off
the (trace of the) generalized Seeley-DeWitt coefficients:
\begin{subequations}
    \label{eq:acoefftrace}
\begin{eqnarray}
{\rm tr}g_0 & = & 2 I_0  \,,
\\
{\rm tr}g_1 & = & 2 a_0^2 \, (I_0 - I_{2t})
\,,
\\
{\rm tr}g_2 & = & {1 \over 3} E^2 \, (I_0 + I_{2t}) + 
a_0^4 \, (I_0 - 2 I_{2t} + I_{4t} )
\,.
\end{eqnarray}
\end{subequations}
These results are in complete agreement with (\ref{eq:acoefftwodim}).

\section{Summary}
\label{sec:sum}

We have presented a generalization of the heat kernel expansion technique to finite
temperature. This task has already been envisaged in \cite{Boschi-Filho:1992ah}. Technically
we followed the approach of \cite{Boschi-Filho:1992ah} to some extent. There are deviations,
however, concerning the formal as well as the physics point of view - which are of
course intertwined. Physically we expect to find (at least) two aspects which make the 
finite temperature case distinct from the vacuum case. First, we expect the appearance
of terms which ``break'' Lorentz invariance like e.g.~a different coefficient
in front of the electric field strength (squared) as compared to the magnetic
field strength. Such differences have not been found in \cite{Boschi-Filho:1992ah} whereas
our calculation shows that they are indeed present (cf.~e.g.~the fifth and sixth
term in (\ref{eq:acoeffcov2})). Second, as already pointed out in the introduction
at finite temperature the field strengths are not sufficient to characterize the
gauge fields. Due to the compactness of the Euclidean time direction the Polyakov
loop carries additional information --- a manifestation of the Aharonov-Bohm effect. 
Hence we expect the appearance of a corresponding
object in the heat kernel expansion. Indeed in our approach the generalized zero mode
which is the logarithm of the Polyakov loop appears. Again this object does not
show up in the approach of \cite{Boschi-Filho:1992ah}. 

Concerning the technical aspects of our approach as compared to 
\cite{Boschi-Filho:1992ah} we have taken care of the correct meaning of the
gauge covariant derivative acting eventually on the arbitrary function $f$.
The latter appears in Nepomechie's constructive method to obtain the heat kernel 
expansion; cf.~(\ref{eq:expanhs}). Only the correct treatment of the covariant
derivative revealed the problem of the naively calculated Seeley-DeWitt coefficients
concerning gauge covariance. Subsequently we developed an expansion method 
which preserves gauge covariance in every single step. While this is not necessary
for the vacuum case --- albeit it does not hurt --- it turned out to be very useful
for the finite temperature case. The requirement to 
simultaneously preserve gauge covariance and (anti-)periodicity led to the introduction
of the generalized zero mode. In this way the Polyakov loop entered our formalism.

{\em Formally} the results of \cite{Boschi-Filho:1992ah} could be reproduced if we replaced 
$I_{2t}$ and $I_{4t}$ by $I_0$ in (\ref{eq:acoeffcov}). Obviously a large number of 
terms would then drop out. 
These terms have not been considered in \cite{Boschi-Filho:1992ah}. By inspecting 
(\ref{eq:i0fin}) it is clear that such a replacement is only correct at zero 
temperature. Consequently the terms which are neglected in \cite{Boschi-Filho:1992ah} are the
ones which are not Lorentz invariant. However, the appearance of such terms is
expected as the heat bath ``breaks'' Lorentz invariance. On the other hand, rotational
invariance is still preserved. Obviously this requirement is fulfilled by our result
(\ref{eq:acoeffcov}). 

Further support for the soundness of our results comes from \cite{Actor:1998cn} where
the case of two-dimensional QED at finite temperature with constant fields is 
addressed. The possible appearance of a zero mode which cannot be gauged away at finite
temperature constitutes the discrepancy between
\cite{Boschi-Filho:1992ah} and \cite{Actor:1998cn} as noted in the latter reference.
As shown above our results are
in complete agreement with the special case of two-dimensional QED with constant fields
discussed in \cite{Actor:1998cn}. One might regard
our results as a generalization of the results of \cite{Actor:1998cn} to arbitrary 
dimensions and non-constant fields. On the other hand, in
contrast to \cite{Actor:1998cn} we cannot offer an exact solution here but only an expansion
in terms of field amplitudes and their derivatives. 

Concerning gauge fields at finite temperature the crucial point is that this expansion 
is not only an 
expansion in powers of the field strength but in addition in powers of the generalized
zero mode. As already emphasized in \cite{Actor:1998cn} even a constant zero mode cannot
be gauged away at finite temperature. It is important to stress one aspect which is
different in our approach as compared to the exact result presented in \cite{Actor:1998cn}.
Inspecting (\ref{eq:actorheat}) reveals a periodicity of the heat kernel with respect
to 
\begin{equation}
  \label{eq:periodzero}
a \to a + 1   \,.
\end{equation}
Our approximate result (\ref{eq:acoefftwodim}) seems not to reproduce that behavior of 
the exact solution. Of course, an expansion in powers of the zero mode amplitude might
not be appropriate any more for large amplitudes. On the other hand, due to the 
periodicity property (\ref{eq:periodzero}) the zero mode can be made arbitrarily large.
Actually this problem is just another manifestation of the fact that the logarithm 
introduced in (\ref{eq:defa0}) is in principle multiple valued. Indeed the Polyakov 
loop which corresponds to the problem studied in \cite{Actor:1998cn} is given by
\begin{equation}
  \label{eq:polyactor}
L(x) = e^{i(\beta E x_1 + 2 \pi a)}   \,.
\end{equation}
Obviously at this stage the periodicity property is still present. The problem appears
with the calculation of the zero mode according to (\ref{eq:defa0}) which apparently
yields (\ref{eq:twodimzeromode}). This, however, is not fully true: If the logarithm
is defined via its Taylor expansion (\ref{eq:deflog}) the outcome of (\ref{eq:defa0}) 
is 
\begin{equation}
  \label{eq:truea0}
a_0(x) = -i \, [E x_1 + 2 \pi T (a-m)] 
\end{equation}
where $m$ is an integer chosen such that $\phi:=\vert \beta E x_1 + 2 \pi (a-m) \vert$
is minimal. And even this is only true if $\phi$ is small enough, 
i.e.~$\vert e^{i\phi}-1 \vert < 1$. 
Otherwise the Taylor expansion (\ref{eq:deflog}) for the logarithm would not converge.
With the correct result (\ref{eq:truea0}) for the zero mode we see that the
periodicity property (\ref{eq:periodzero}) is still present. On the other hand,
by considering the convergence properties of the Taylor expansion for the logarithm 
it becomes obvious 
that for an expansion of the heat kernel small amplitudes are mandatory
right from the start. Strictly speaking so far our formalism can only be applied to
Polyakov loops with values close to unity. Of course, nothing is wrong with an
expansion in powers of the field amplitudes which works when the amplitudes are small
but breaks down when they are large. On the other hand, it clearly would be
desirable to develop a heat
kernel expansion in which the Polyakov loop enters directly and not only via the
generalized zero mode. At the present stage it is not clear whether this is possible.

\appendix

\section{Useful sum-integrals}
\label{sec:app}

The following sum-integrals show up in the calculation of the lowest order terms of the
heat kernel expansion:
\begin{subequations}
  \label{eq:sumint}
\begin{eqnarray}
(4 \pi \tau)^{D/2} \sumint {d^{D}k \over (2 \pi)^{D}} \, e^{-\tau k^2} & =: & 
I_0(\sqrt{\tau}T)   \,,  \\
(4 \pi \tau)^{D/2} \sumint {d^{D}k \over (2 \pi)^{D}} \, e^{-\tau k^2} \, 2 \tau k_0^2
& =: & I_{2t}(\sqrt{\tau}T)   \,,  \\
(4 \pi \tau)^{D/2} \sumint {d^{D}k \over (2 \pi)^{D}} \, e^{-\tau k^2} \, 
2 \tau k_i k_j
& =: & I_{2s}(\sqrt{\tau}T) \, \delta_{ij}   \,,  \\
(4 \pi \tau)^{D/2} \sumint {d^{D}k \over (2 \pi)^{D}} \, e^{-\tau k^2} \, 
{4\over 3} \tau^2 k_0^4
& =: & I_{4t}(\sqrt{\tau}T)   \,,  \\
(4 \pi \tau)^{D/2} \sumint {d^{D}k \over (2 \pi)^{D}} \, e^{-\tau k^2} \, 
4 \tau^2 k_0^2 k_i k_j
& =: & I_{4m}(\sqrt{\tau}T) \, \delta_{ij}   \,,  \\
(4 \pi \tau)^{D/2} \sumint {d^{D}k \over (2 \pi)^{D}} \, e^{-\tau k^2} \, 
4 \tau^2 k_i k_j k_l k_m
& =: & I_{4s}(\sqrt{\tau}T) \, 
\left(\delta_{ij} \delta_{lm} + \delta_{il} \delta_{jm} + \delta_{im} \delta_{jl} \right)
\,.
\end{eqnarray}
\end{subequations}
Note that sum-integrals with an odd number of $k$ factors vanish. The spatial 
$k$-integrations can be readily evaluated using
\begin{subequations}
  \label{eq:spaceint}
\begin{eqnarray}
\int {d^{D}k \over \pi^{D}} \, e^{-\vec k^2} & = & 1 \,  \\ 
\int {d^{D}k \over \pi^{D}} \, e^{-\vec k^2} \, k_i k_j & = & 
{1 \over 2} \, \delta_{ij}   \,,  \\
\int {d^{D}k \over \pi^{D}} \, e^{-\vec k^2} \, k_i k_j k_l k_m & = & {1 \over 4}
\left(\delta_{ij} \delta_{lm} + \delta_{il} \delta_{jm} + \delta_{im} \delta_{jl} \right)
\,.
\end{eqnarray}
\end{subequations}
One gets
\begin{subequations}
  \label{eq:timesum}
\begin{eqnarray}
I_0(\sqrt{\tau}T) = I_{2s}(\sqrt{\tau}T) = I_{4s}(\sqrt{\tau}T) & = & 
(4 \pi \tau)^{1/2} T \sum\limits_{n = -\infty}^{+\infty} \exp(-\tau \omega_n^2)  \,, \\
I_{2t}(\sqrt{\tau}T) = I_{4m}(\sqrt{\tau}T) & = & 
(4 \pi \tau)^{1/2} T \sum\limits_{n = -\infty}^{+\infty} \exp(-\tau \omega_n^2)  \,
2 \tau \omega_n^2  \,, \\
I_{4t}(\sqrt{\tau}T) & = & 
(4 \pi \tau)^{1/2} T \sum\limits_{n = -\infty}^{+\infty} \exp(-\tau \omega_n^2)  \,
{4 \over 3} \tau ^2 \omega_n^4  \,.
\end{eqnarray}
\end{subequations}
These finite-temperature sums can be rewritten using a generalized theta function
transformation. We adopt formula (3.11) from \cite{Boschi-Filho:1992ah} (with a misprint corrected):
\begin{equation}
  \label{eq:gentheta}
\sum\limits_{n = -\infty}^{+\infty} \exp\left( -\gamma\,(n+1/2)^2 \right) \,
\left[ \gamma \, (n+1/2)^2 \right]^p 
= \left( {\pi \over \gamma} \right)^{1/2} \, {(2p-1)!! \over 2^p}
\left[ 1 + 2 \sum\limits_{n=1}^\infty (-1)^n \, \exp(-\pi^2 n^2/\gamma)
\sum\limits_{l=0}^p \left( {-\pi^2 n^2 \over \gamma} \right)^l C_l^p \right]
\end{equation}
with
\begin{equation}
  \label{eq:defclp}
C_l^p := { 2^l \, p! \over l! \, (p-l)! \, (2l-1)!! }  \,,
\end{equation}
$p$ a non-negative integer and $(2p-1)!! := 1 \cdot 3 \cdot \ldots \cdot (2p-1)$.
We apply (\ref{eq:gentheta}) to (\ref{eq:timesum}) with $\gamma = (2\pi T)^2 \tau$:
\begin{subequations}
  \label{eq:i0fin}
\begin{eqnarray}
I_0 & = & 1 + 2 \sum\limits_{n=1}^\infty (-1)^n \, 
\exp\left(-{n^2 \over 4 \tau T^2} \right)   \,,
\\ 
I_{2t} & = & 1 + 2 \sum\limits_{n=1}^\infty (-1)^n \, 
\exp\left(-{n^2 \over 4 \tau T^2} \right) \,
\left( 1 -2 \, {n^2 \over 4 \tau T^2} \right)  \,,
\\
I_{4t} & = & 1 + 2 \sum\limits_{n=1}^\infty (-1)^n \, 
\exp\left(-{n^2 \over 4 \tau T^2} \right) \,
\left[ 1 - 4 \, {n^2 \over 4 \tau T^2} +  {4 \over 3} 
\left( {n^2 \over 4 \tau T^2} \right)^2 \right]  \,.
\end{eqnarray}
\end{subequations}
We also present some combinations of $I_{\dots}$-functions which appear in 
(\ref{eq:acoeffcov}):
\begin{subequations}
  \label{eq:i0combi}
\begin{eqnarray}
I_0 - I_{2t} & = & \sum\limits_{n=1}^\infty (-1)^n \, 
\exp\left(-{n^2 \over 4 \tau T^2} \right) \, {n^2 \over \tau T^2}   \,,
\\ 
I_0 - 2 I_{2t} + I_{4t} & = & {1 \over 6} \sum\limits_{n=1}^\infty (-1)^n \, 
\exp\left(-{n^2 \over 4 \tau T^2} \right) \,
\left( {n^2 \over \tau T^2} \right)^2   \,.
\end{eqnarray}
\end{subequations}

\section{Properties of the generalized zero mode}
\label{sec:a0prop}

In this appendix we collect and prove some important properties for the generalized
zero mode defined in (\ref{eq:defa0}). We first note that for an abelian gauge theory
$a_0$ is just the zero mode of the Fourier decomposition of the time component of the
vector potential:
\begin{equation}
  \label{eq:a0abel}
A_0(x_0,\vec x) = 
\sum\limits_{n=-\infty}^{+\infty} a_n(\vec x) \, e^{i 2 n \pi T x_0} \,.
\end{equation}
For a non-abelian gauge theory such an easy relation cannot be obtained. The abelian
case, however, justifies our name choice ``generalized zero mode''. 
The relations presented in the following hold for arbitrary gauge theories. 

Periodicity:
\begin{equation}
  \label{eq:a0period}
a_0(x_0+\beta,\vec x) = -T \log P \exp 
\left[-\int\limits_{x_0+\beta}^{x_0+2\beta} dt \, A_0(t,\vec x) \right] 
= -T \log P \exp 
\left[-\int\limits_{x_0}^{x_0+\beta} dt \, 
\underbrace{A_0(t+\beta,\vec x)}_{=A_0(t,\vec x)} \right] 
= a_0(x_0,\vec x)   \,.
\end{equation}

Behavior under gauge transformations: \\
We restrict ourselves to gauge transformations which are periodic in $x_0$. These are
the only ones appropriate for the fermion fields which are integrated out via the
heat kernel method. On account of (\ref{eq:Utrafo}) we find
\begin{equation}
  \label{eq:poltrafo}
U[A](x_0+\beta,\vec x;x_0,\vec x) \to 
V(x_0+\beta,\vec x) \, U[A](x_0+\beta,\vec x;x_0,\vec x) \, V^{-1}(x_0,\vec x)
= V(x_0,\vec x) \, U[A](x_0+\beta,\vec x;x_0,\vec x) \, V^{-1}(x_0,\vec x)   \,.
\end{equation}
To obtain the transformation property of $a_0$ we decompose $U$ as
\begin{equation}
  \label{eq:defcalO}
U[A](x_0+\beta,\vec x;x_0,\vec x) =: 1 + {\cal O}
\end{equation}
and use a Taylor expansion of the log-function:
\begin{eqnarray}
  \label{eq:taylorlog}
\log U[A](x_0+\beta,\vec x;x_0,\vec x) & \to & \log( V U V^{-1}) = 
\log( V \, (1 + {\cal O}) V^{-1}) = \log(1+ V {\cal O} V^{-1}) = 
\sum\limits_{n=1}^\infty (-1)^{n+1} \, {1 \over n} \, (V {\cal O} V^{-1})^n 
\nonumber \\ && =
V \log(1 + {\cal O}) V^{-1} = V \, (\log U) \, V^{-1}  \,.
\end{eqnarray}
Thus
\begin{equation}
  \label{eq:a0trafo}
a_0(x) \to V(x) \, a_0(x) \, V^{-1}(x)   \,.
\end{equation}
Finally one can also show using the Taylor expansion of the exponential function:
\begin{equation}
  \label{eq:trafoexp}
\exp\left(\alpha a_0(x)\right) \to \exp\left(\alpha V(x) \, a_0(x) \, V^{-1}(x)\right) = 
V(x) \, \exp\left(\alpha a_0(x) \right) \, V^{-1}(x)
\end{equation}
for arbitrary numbers $\alpha$.

\section{Proof of equation (\ref{eq:timeder})}
\label{sec:proof}

In \cite{Elze:1986qd} the derivative of a straight line gauge link $U_S[A](b,a)$ 
with respect to its 
endpoints $a(x)$, $b(x)$ has been considered. One gets
\begin{eqnarray}
  \label{eq:elzeder}
\partial_\mu^x U_S[A](b,a) & = & - {\partial b_\nu \over \partial x_\mu}
\left[ A_\nu(b) - (b-a)_\alpha \int\limits_0^1 \!\! ds \, s \, U_S[A](b,z) \,
F_{\alpha\nu}(z) \, U_S[A](z,b) \right] \, U_S[A](b,a)
\nonumber \\ && {}
+ U_S[A](b,a) \, \left[ A_\nu(a) - (b-a)_\alpha \int\limits_0^1 \! \! ds \, (s-1) \,
U_S[A](a,z) \, F_{\alpha \nu}(z) \, U_S[A](z,a) \right] 
{\partial a_\nu \over \partial x_\mu}
\end{eqnarray}
with the straight line path connecting $a$ and $b$,
\begin{equation}
  \label{eq:defz}
z = z(s) = a + s \, (b-a)  \,,
\end{equation}
and the field strength
\begin{equation}
  \label{eq:deffieldstrength}
F_{\mu\nu} := [D_\mu,D_\nu]  \,.
\end{equation}
This implies for the gauge links of interest
\begin{equation}
  \label{eq:derivU}
\left[ D^x_0,U[A](x_0+\beta,\vec x;x_0,\vec x) \right] = 0
\end{equation}
and
\begin{equation}
  \label{eq:derivU2}
D^x_0 \, U[A](x;x') = U[A](x;x') \, \partial_0^x   \,.
\end{equation}
We make use of the decomposition of $U$ given in (\ref{eq:defcalO}). 
Using (\ref{eq:derivU}) we find
\begin{equation}
  \label{eq:commO}
\left[ D_0,{\cal O} \right] = 0
\end{equation}
and therefore
\begin{equation}
  \label{eq:commOn}
\left[ D_0,{\cal O}^n \right] = 0   \,.
\end{equation}
This yields
\begin{equation}
  \label{eq:comma0}
[D_0,a_0] = -T \, [D_0,\log(1+{\cal O})] 
= -T \sum\limits_{n=1}^\infty (-1)^{n+1} \, {1 \over n} \, [D_0,{\cal O}^n] = 0  \,.
\end{equation}
Hence we obtain
\begin{eqnarray}
\left[ D_0,\exp\left(x_0 \,a_0(x)\right) \right] & = & 
\sum\limits_{n=0}^\infty {1 \over n!} \, \left[D_0,x_0^n \, a_0^n \right]
= \sum\limits_{n=0}^\infty {1 \over n!} \, \left[D_0,x_0^n \right] \, a_0^n 
+ \sum\limits_{n=0}^\infty {1 \over n!} \, x_0^n \, \left[D_0,a_0^n \right]
= \sum\limits_{n=0}^\infty {1 \over n!} \, n x_0^{n-1} \, a_0^n 
\nonumber \\
  \label{eq:D0exp}
& = & a_0 \, \exp\left(x_0 \,a_0(x)\right)   \,.
\end{eqnarray}
Finally we get
\begin{eqnarray}
D_0^x \tilde U[A](x;x') & = &
D_0^x e^{x_0 \, a_0(x)} \, U[A](x;x') \, e^{-x_0' \, a_0(x')} 
\nonumber \\
& = & \left[D_0^x , e^{x_0 \, a_0(x)} \right] \, U[A](x;x') \, e^{-x_0' \, a_0(x')} 
+ e^{x_0 \, a_0(x)} D_0^x \, U[A](x;x') \, e^{-x_0' \, a_0(x')} 
\nonumber \\
  \label{eq:finproof}
& = & a_0(x) \, \tilde U[A](x;x') 
\end{eqnarray}
where we have used (\ref{eq:derivU2}) in the last step. Using (\ref{eq:finproof}),
(\ref{eq:comma0}) and (\ref{eq:propUtilde1}) leads to (\ref{eq:timeder}).

\bibliography{literature}
\bibliographystyle{apsrev}

\end{document}